\documentclass[twocolumn]{aa}
\usepackage{graphicx}
\usepackage{natbib}
\bibpunct{(}{)}{;}{a}{}{,}
\usepackage{txfonts}
\begin{document}
\title{Scale sensitive deconvolution of interferometric images}
\subtitle{I. Adaptive Scale Pixel (Asp) decomposition}

\author{S. Bhatnagar \inst{1,2} \and T.J. Cornwell \inst{1,3}}

\offprints{S. Bhatnagar}

\institute{National Radio Astronomy Observatory\thanks{Associated
Universities Inc. operates the National Radio Astronomy Observatory
under cooperative agreement with the National Science Foundation},
           1003 Lopezville Road, Socorro, NM-87801, USA.
           \and
           \email{sbhatnag@aoc.nrao.edu} \and
           \email{tcornwel@aoc.nrao.edu}
}
\date{Received: Feb., 2004; revised: May, 2004; accepted date}

\abstract{
Deconvolution of the telescope Point Spread Function (PSF) is
necessary for even moderate dynamic range imaging with interferometric
telescopes.  The process of deconvolution can be treated as a search
for a model image such that the residual image is consistent with the
noise model.  For any search algorithm, a parameterized function
representing the model such that it fundamentally separates signal
from noise will give optimal results.  In this paper, the first in a
series of forthcoming papers, we argue that in general, spatial
correlation length (a measure of the scale of emission) is a stronger
separator of the signal from the noise, compared to the strength of
the signal alone.  Consequently scale sensitive deconvolution
algorithms result into more noise-like residuals.  We present a
scale-sensitive deconvolution algorithm for radio interferometric
images, which models the image as a collection of Adaptive Scale
Pixels (Asp).  Some attempts at optimizing the runtime performance are
also presented.
 }  

\maketitle
%

\section{Introduction}

Interferometric telescopes employ the van Cittert-Zernike theorem to
synthesize apertures much larger than the size of the individual
antennas \citep{THOMPSON_AND_MORAN}.  Such instruments measure the
source coherence function, which is the Fourier transform of the sky
brightness distribution.  However the source coherence function is
measured at discrete points in the Fourier space resulting into a
point spread function (PSF) which has significant wide-spread
sidelobes.  The presence of these sidelobes limits the dynamic range
of raw images made with such telescopes.  For even moderate dynamic
range imaging, deconvolution of the PSF is necessary.

The measurement equation describing an interferometer can be written
as:
\begin{equation}
\label{ME}
 \vec{V} = \tens{A} \vec{I^o} + \tens{A} \vec{N}
\end{equation}
where $\vec{V}$ and $\vec{N}$ are the measured visibility and
independent random noise vectors respectively, and $\vec{I^o}$ is the
true image.  The measurement matrix $\tens{A}$ represents the linear
transform from the image to the visibility domain.  For perfectly
calibrated co-planar Fourier synthesis devices like imaging
interferometric telescopes, $\tens{A}$ is the rectangular
observation matrix, the non-zero elements of which are the sines and
cosines corresponding to the measured fourier components.  In
practice, due to incomplete sampling of the visibility domain,
$\tens{A}$ in general is a singular non-square matrix, excluding the
use of linear methods to invert the above equation.  Non-linear
deconvolution methods are therefore used to estimate $\vec{I^o}$.  In
the usual terminology used in the literature, $\tens{A}^{\rm T}
\vec{V}$ is the dirty image ($\vec{I^d}$) and $\tens{A}^{\rm T}
\tens{A}=\tens{B}$ is the beam matrix (the super-script $\rm T$
implies a transpose).  $\tens{B}$ is a toeplitz matrix of the dirty
beam or the point spread function and $\tens{B}\vec{I^o}$ represents
the convolution of the true image with the PSF.  Note that the
resultant noise vector in the image domain ($\tens{A}^{\rm T} \tens{A}
\vec{N}$) is also convolved with the PSF and the pixel-to-pixel noise
in the image is not independent.

The process of deconvolution can be described as a search for a model
image $\vec{I^M}$ which solves the normal equation
\begin{equation}
\label{NORMEQ}
\tens{A}^{\rm T} \vec{V} = \tens{A}^{\rm T} \tens{A} \vec{I^M} + \tens{A}^{\rm T} \tens{A}\vec{N}
\end{equation}
A non-linear minimization scheme can be set up which iteratively
minimizes the objective function until the residuals
$\vec{V}-\tens{A}\vec{I^M}$ are noise-like.  

The dimensionality and the nature of the search space is governed by
the parameterization of $\vec{I^M}$.  Scale insensitive deconvolution
algorithms like CLEAN
\citep{CLEAN} and its variants model the image as
\begin{equation}
\label{MODIMG}
\vec{I^M}=\sum_k F_k\delta(x-x_k,y-y_k)
\end{equation}
which is a collection of delta functions of amplitude $F_k$ at each
pixel location.  An $N\times M$ pixel clean-box corresponds to a
$N\times M$ dimensional search space.  The CLEAN algorithm iteratively
estimates the $F_k$s by taking a fixed sized step along the axis of
highest derivative (the peak in the residual image) \citep[see][for a
more complete description]{SCHWARZ_CLEAN}.  This is done by updating
the residual image at each iteration as $\vec{I}^{\vec{R}}_k =
\vec{I}^{\vec{R}}_{k-1} - g\tens{B}\left[\max{\vec{I}^{\vec{R}}_{k-1}}\right]$
where $g$ is called the loop-gain (it controls the step-size).  This
implicitly assumes an orthogonal search space of constant curvature.
Variants of CLEAN \citep{CLARK_CLEAN} operate in two cycles called
major and minor cycles.  The minor cycle uses an approximate PSF and
builds a shallow model image.  The accuracy lost in the minor cycle
due to the use of approximate PSF is recovered in the more expensive
major cycle where the residual image is computed at full accuracy as
$\tens{A}^{\rm T}\left[\vec{V}-\tens{A} \vec{I^M}\right]$.  It can
therefore be considered to be a steepest descent algorithm to minimize
the objective function
$\chi^2=\left[\vec{V}-\tens{A}\vec{I^M}\right]^{\rm T} \tens{W}
\left[\vec{V}-\tens{A}\vec{I^M}\right]$.  The dimensionality of the
search space can be constrained by the user defined clean-box which
usually remains fixed after initial specification.  The step size is
also a user defined parameter, typically $\le 0.2$.  The stopping
criterion is a combination of the maximum number of iterations and the
magnitude of the maximum residual.  Typically, the algorithm is
stopped (often by user intervention or by adjusting the maximum number
of iterations) when maximum residuals are comparable to the estimated
noise (another user defined parameter).  The regularization to avoid
the problem of over fitting in such an unconstrained minimization is
explicitly imposed via forcing a finite number of iterations.

Statistical image reconstruction algorithms like Maximum Entropy
Method (MEM) \citep[][and references therein]{MEM_ARAA} on the other
hand set up a formal constrained minimization algorithm which
minimizes the objective function $f(\vec{I^M},\lambda) =
H-\lambda\chi^2$ where $H$ is the entropy
function, $\lambda$ is a Lagrange multiplier and the model image is
parameterized as in Eq.~\ref{MODIMG}.  The function $H$ is
derived from a physically meaningful prior distribution and imposes
various desirable constraints like smoothness.  An extra term,
sometimes with another undetermined Lagrange multiplier is also used
to impose the positivity constraint.  The Entropy function acts as a
regularizer, biasing the solution towards the supplied prior image
while $\chi^2$ pulls the solution towards the best fit to the data.
From a set of images all of which satisfy the normal equation
(Eq.~\ref{NORMEQ}), the MEM image corresponds to the mode of the
a-posterior distribution.  The step size computation for the
minimization of $f$ requires an evaluation of the inverse of the
Hessian matrix.  For images with a large number of pixels with
significant emission, the size of the Hessian matrix can be large and
inverting it typically requires SVD like algorithms which are
computationally expensive.  \citet{MEM} devised a fast iterative
algorithm by approximating the Hessian by a diagonal matrix to gain in
speed.

Both approaches however use the parameterization in Eq.~\ref{MODIMG}
which we refer to as a scale-less model.  The image is decomposed into
a set of delta functions and the search space is assumed to be
orthogonal.

Coupling of the pixels in the dirty image comes from two sources.  The
pixels of the underlying true image are inherently not independent in
the presence of extended emission.  The second source of coupling
comes from the PSF.  As mentioned earlier, the main lobe of the PSF
has a finite width (proportional to the diffraction limit of the
synthesized aperture) and significant wide-spread sidelobes.  The
former kind of coupling implies that the search space is potentially
non-orthogonal unless the true sky is composed only of clearly
unresolved sources.  The latter however implies a coupling of even
widely separated unresolved sources via the sidelobes of the PSF.
Ignoring the coupling due to the PSF possibly only results into slower
convergence.  However, ignoring the inherent coupling of pixels for
extended emission allows more degrees of freedom (DOF) than required,
which results into the breaking up of the extended emission leaving
low level correlated residuals.  The peak residuals are comparable to
the estimated noise but are correlated at scales larger than the
resolution element (the synthesized beam).  In the absence of
any scale information in Eq.~\ref{MODIMG}, such residual emission
cannot be recovered.

It is easy to see that an optimal image reconstruction algorithm will
also use the minimum number of DOF to represent the image.  However in
practice, it is almost impossible to determine this minimum number,
and hence also difficult to design an algorithm which will achieve
such a representation Practically therefore, the goal is to seek a
solution with least complexity.  From the point of view of imaging
weak extended emission, the important improvement in the deconvolution
algorithm is therefore to decompose the true sky image in a scale
sensitive basis.  This implies a significant increase in algorithm
complexity and computational cost.  In this paper we describe a scale
sensitive deconvolution algorithm for interferometric images and
attempts at improving its performance.
\section{Fundamental separation of signal and noise}

Equation~\ref{NORMEQ} can be written in terms of the dirty and true
images as
\begin{equation}
\label{NORMEQ_IM}
\vec{I^d} = \tens{B}\vec{I^o} + \vec{I^N}~~~~{\rm
where}~~~\vec{I^N}=\tens{B}\vec{N} 
\end{equation}
The distinction between the true and the dirty image can then be
stated as the latter being equal to the true image corrupted in a
deterministic way by the sidelobes of the PSF and in a
non-deterministic way by the additive noise image ($\vec{I^N}$).
These corruptions are represented by the two terms in
Eq.~\ref{NORMEQ_IM}.  Prior knowledge of the deterministic pattern of
the PSF is explicitly used by deconvolution algorithms which attempt
to recover $\vec{I^o}$ from $\vec{I^d}$ given $\tens{B}$.  This
approach works well where the first term in the above equation
dominates.  For weak large scale emission, the effect of the second
term is significant.  It is easy to see that without explicitly
incorporating prior information which fundamentally separates the two
terms, such features in $\vec{I^o}$ cannot be recovered.  Indeed, it
is well known that the residuals of scale-less deconvolution are
correlated with the large scale features and large scale features are
in general poorly reconstructed.

The term involving $\vec{N}$ in Eq.~\ref{ME} can be shown to be a
gaussian random process \citep{THOMPSON_AND_MORAN}, with the
pixel-to-pixel noise being independent (zero correlation length).
Ideally, in the absence of the convolution with the PSF, this would
result into the noise image $\vec{I^N}$ with a auto-correlation
function (ACF) of zero width.  However the main lobe of the PSF (the
synthesized beam or the nominal resolution element) has a width larger
than the image pixel size.  Convolution by the PSF therefore results
into smoothing of all pixel-to-pixel variations at scales smaller than
the synthesized beam making the auto-correlation width of $\vec{I^N}$
of the order of the resolution element.  This implies that the {\it
largest} correlated feature in $\vec{I^N}$ is of the order of the
resolution element.  On the other hand any physically plausible model
image representing an observation of $\vec{I^o}$ will have {\it
minimum} correlation length of the order of the resolution element of
the imaging device.  The range of allowed scales of emission
(correlation length) therefore provides a fundamental handle for
separating $\vec{I^N}$ from $\vec{I^d}$.  Correlated emission in
$\vec{I^d}$ at scales much larger than the resolution element must
come from real extended emission.  Correlation length alone for such
emission fundamentally separates it from the noise.  For unresolved
features (scale comparable to the resolution element) in $\vec{I^o}$,
the strength of the emission provides a similar handle (features at
scales smaller than, or equal to the resolution element and weaker
than the estimated noise RMS are indistinguishable from $\vec{I^N}$).
The ACF of $\vec{I^N}$, could have low level wings at scales
marginally larger than the main lobe (due to the sidelobes of the
PSF).  However, for observations with good uv-coverage\footnote{Good
uv-coverage is a pre-requisite for high fidelity and dynamic range
imaging of extended emission.}, the sidelobes are small and the
resulting wings in the ACF will be at a very low level.  For real
emission at scales in this regime where the fundamental scales of the
noise in the image domain and the real emission overlap, the strength
of emission plays a crucial role along with the scale of emission in
separating the signal from the noise (correlated emission at these
scales, which is also significantly stronger than the expected noise
is more likely to be due to the real emission).  A combination of
scale and the strength of emission therefore fundamentally separates
noise from the signal.  Scale sensitive deconvolution algorithms
incorporate this information explicitly in the deconvolution process
and hence leave more noise-like residuals.

\section{Overview of scale sensitive methods}

Scale sensitive (multi-scale) methods seek to represent the inherent
coupling of the pixels in the true image by decomposing it in a basis
which locally minimizes the complexity of representation.  Since the
use of minimum DOF also corresponds to minimum complexity, in practice
the problem reduces to finding the largest {\it locally} best fit
scale in the image.  Clearly, the efficiency with which this can be
done critically depends on the degree of coupling between pixels due
to the PSF.  The larger the scale of this coupling, the more complex
is the structure of the covariance matrix.  Consequently, the
deconvolution algorithms become more complex and computationally
expensive.  Efficient scale sensitive deconvolution algorithms for
images from filled aperture instruments with additive {\it
uncorrelated} noise now exist.  Efficient algorithms for the
deconvolution of interferometric images, where the PSF has significant
large scale sidelobes and the noise in the image domain is correlated,
pose a bigger challenge in comparison to filled aperture telescopes.

\subsection{The Pixon method: for filled aperture instruments}

The Pixon method \citep{PIXON0} iteratively decomposes the true image
as a collection of locally best-fit kernels (usually gaussians).  A
kernel is used for every pixel in the image and the parameters of
these kernels are determined by a localized fitting to the data using
a user defined goodness of fit (GOF) criteria.  A kernel is allowed to
be as wide as possible while simultaneously satisfying the GOF
locally.  For the case of filled aperture telescopes, the PSF has a
limited support and along with the local kernel, provides a limited
footprint on the data (the raw image) making the concept of ``local''
well defined.  As one can imagine, as the decomposition proceeds, many
kernels will cease to be significant (corresponding to pixels which
are better represented as part of a larger kernel).  A patented fast
Pixon algorithm exists \citep{PIXON_ADASS}, the details of which are
unfortunately not known due to patent restrictions.  Since the
criteria for the acceptance of the kernels is enforced only locally,
the method can deal with non-uniform noise across the image and
recovers low level large scale features effectively.

The assumptions of a finite support PSF and additive {\it independent}
noise in the images are central to this method.  Therefore, despite
its impressive successes with images from a wide variety of filled
aperture imaging devices, it is not suited for interferometric imaging
where both these assumptions are grossly invalid.

\subsection{The Multi-scale Clean: for interferometric instruments}

The Multi-scale Clean (MS-Clean) method \citep{MS_CLEAN} is motivated
by the Pixon approach and the usual CLEAN algorithm.  The update
direction at each iteration is given by the residual image which
involves a convolution of the current model image with the PSF.  In
general, because of the complex structure of the PSF, this convolution
has to be numerically computed.  The normal CLEAN algorithm gains in
performance by modeling the emission as in Eq.~\ref{MODIMG} where the
convolution with the PSF reduces to a scale-shift-and-add (of the PSF)
operation.  MS-Clean retains this scale-shift-and-add nature of the
algorithm by modeling the emission as a collection of symmetric
gaussians at a few scales.  The convolution of these components with
the PSF is pre-computed.  Versions of the current residual image
smoothed by these gaussians are also maintained.  At each iteration, a
global peak among all these enumerated scales is searched and the
version of the pre-computed convolved PSF at the scale at which the
peak was found is subtracted from the residual images at all scales.
A gaussian of the scale at which the peak was found is added to the
list of components.

This algorithm better recovers the large scale emission than does the
scale-less CLEAN algorithm.  However the scale sizes are restricted to
the few enumerated scales.  Secondly, since the PSF at various scales
are pre-computed, non-symmetric components instead of circularly
symmetric gaussians are expensive to use.  As a result, non-symmetric
features are broken up into a series of smaller scale components.
This leads to the same problem of breaking of structure -- only the
error is at lower spatial frequencies.  Also, since the removal of
components at each successive iteration is decoupled from all previous
components, errors in earlier iterations can only be compensated by
adding more components such that the errors are corrected.  In the
space of the hyper-parameters (the enumerated scales), it effectively
retains the assumption of an orthogonal search space (diagonal
approximation of the Hessian).

\section{The Asp image reconstruction algorithm}

The general problem of scale-sensitive deconvolution is that of a
function optimization in a high dimensional, non-orthogonal space.
The curvature and dimensionality of the space is largely determined by
the parameterization of $\vec{I^M}$ and the extent of the PSF.  The
Pixon method exploits the locality of the effects of the PSF to limit
the dimensionality of the search space.  On the other hand, MS-Clean
explicitly limits the dimensionality of the space by decomposing
$\vec{I^M}$ into a fixed set of few scales.  This fixed set is
determined before hand which remains unchanged from iteration to
iteration and no other scale other than those in this set is
admissible. The Adaptive Scale Pixel (Asp) decomposition method
estimates the best fit Asp at the location of the peak in the residual
image at each iteration.  Due to the inherent coupling of pixels in
the true image as well as due to the extent of the PSF, typically only
a sub-set of the Aspen change significantly at each iteration.  We
refer to this sub-set as the ``active-set''.  This active-set of Aspen
is determined at each iteration.  The number of Aspen in this set
determines the dimensionality of the search space at each iteration.
However, since the members of the active-set are determined
on-the-fly, the set of scales used at each iteration potentially
changes from iteration to iteration as well as all possible scales are
admissible.  This is in contrast with the MS-Clean where only the
selected set of scales are allowed and the set of scales remain fixed
for all iteration.  This effectively relaxes the MS-Clean assumption
of an orthogonal space (i.e., the Aspen estimated in earlier
iterations are subject to change in later iterations) and allows
errors in earlier iterations to be compensated, to the extent
possible, by adjusting the current set of active Aspen.  This
potentially also reduces the total number of components required.

\begin{figure*}
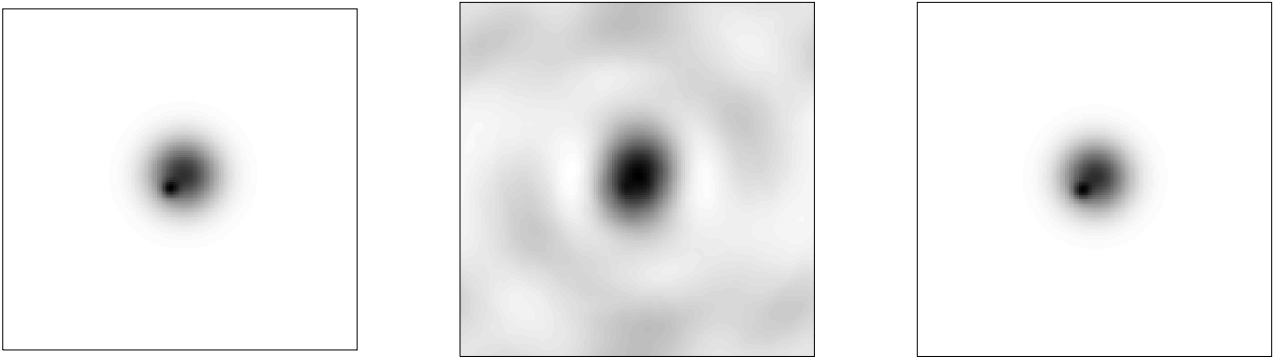

  \centering 
\hbox{
	\includegraphics[width=6cm]{overlap.orig.eps}
	\includegraphics[width=6cm]{overlap.di.eps}
	\includegraphics[width=6cm]{overlap.ci.eps}
}

  \caption{This figure shows an example using simulated image
  composed of two components with significantly different but
  overlapping scales (left).  A dirty image (center) was formed by the
  convolution of this model image with a typical interferometric PSF
  and deconvolved using the Asp-Clean algorithm.  The Asp model image
  is shown in the right panel.  Only two significant components were
  required for the residuals to be noise-like.} \label{FigExample}
\end{figure*}

The general Asp decomposition of the image is expressed as:
\begin{equation}
\vec{I^M} = \sum_k \vec{P}(p_k)
\end{equation}
where $p_k \equiv\{Amplitude, Location, Scale\}$.  A simple functional
form for $\vec{P}$ can be a circularly symmetric gaussian or any other
more appropriate form.  The best fit Aspen are found by minimizing the
objective function
\begin{equation}
\chi^2 = \left[\vec{V} - \tens{A}\vec{I^M}\right]^{\rm T} \tens{W} \left[\vec{V}-\tens{A}\vec{I^M}\right]
\end{equation}
where $\vec{I^M}$ is the current model image and $\tens{W}$ is the
weights matrix.  The update direction computation requires evaluation
of the variation of $\chi^2$ with respect to $p_k$s given by
\begin{equation}
\frac{\partial\chi^2}{\partial p_k} = -2 \left[\vec{I^R}\right]^{\rm T} \vec{\frac{\partial
P}{\partial p_k}}~~~~~~~{\rm where}~~\vec{I^R} = \vec{I^d} - \tens{B}\vec{I^M}
\end{equation}

The parameters corresponding to the location, amplitude and scale are
estimated at each iteration by searching for a peak in the residual
images smoothed to a few scales.  In practice, $\chi^2$ was found to
be weakly dependent on the change in the position of the Aspen.  The
location of the Aspen is therefore kept fixed at the location of the
peak, while the amplitude and scale parameters are adjusted by
fitting.  A new Asp is added at each iteration unless one of the
termination criteria is satisfied.  The various steps are therefore:

\begin{enumerate}

\item Set $\vec{I}^{\vec{R}}_{\vec{o}}=\vec{I^d}$.

\item \label{START} At the $n^{th}$ iteration, smooth
$\vec{I}^{\vec{R}}_n$ by a gaussian beam at a few scales from the
resolution element to a few times the resolution element.

\item Search for a global peak ($F_k$) among these smoothed residual
images. Define a new Asp $\vec{A}_k$ with amplitude $F_k$ centered at
the location of the peak $(x_k,y_k)$ with scale $\sigma_k$.  Add it to
the current set of Aspen $\{\vec{P}\}_n=\{\vec{P}\}_{n-1} \cup
\{\vec{A}_k\}$.

\item \label{OPT} Find the best-fit set $\{\vec{P}\}_n$.

\item Compute $\vec{I}^{\vec{M}}_n$.  

\item Update the residual image as
$\vec{I}^{\vec{R}}_{n+1}=\vec{I}^{\vec{d}}-\tens{B}\vec{I}^{\vec{M}}_n$.

\item Goto step~\ref{START} unless the termination criteria is met or
the residuals are noise-like.

\end{enumerate}
\begin{figure*}
\begin{center}
   \vbox{ 
        \hbox{ 
                \includegraphics[width=8cm]{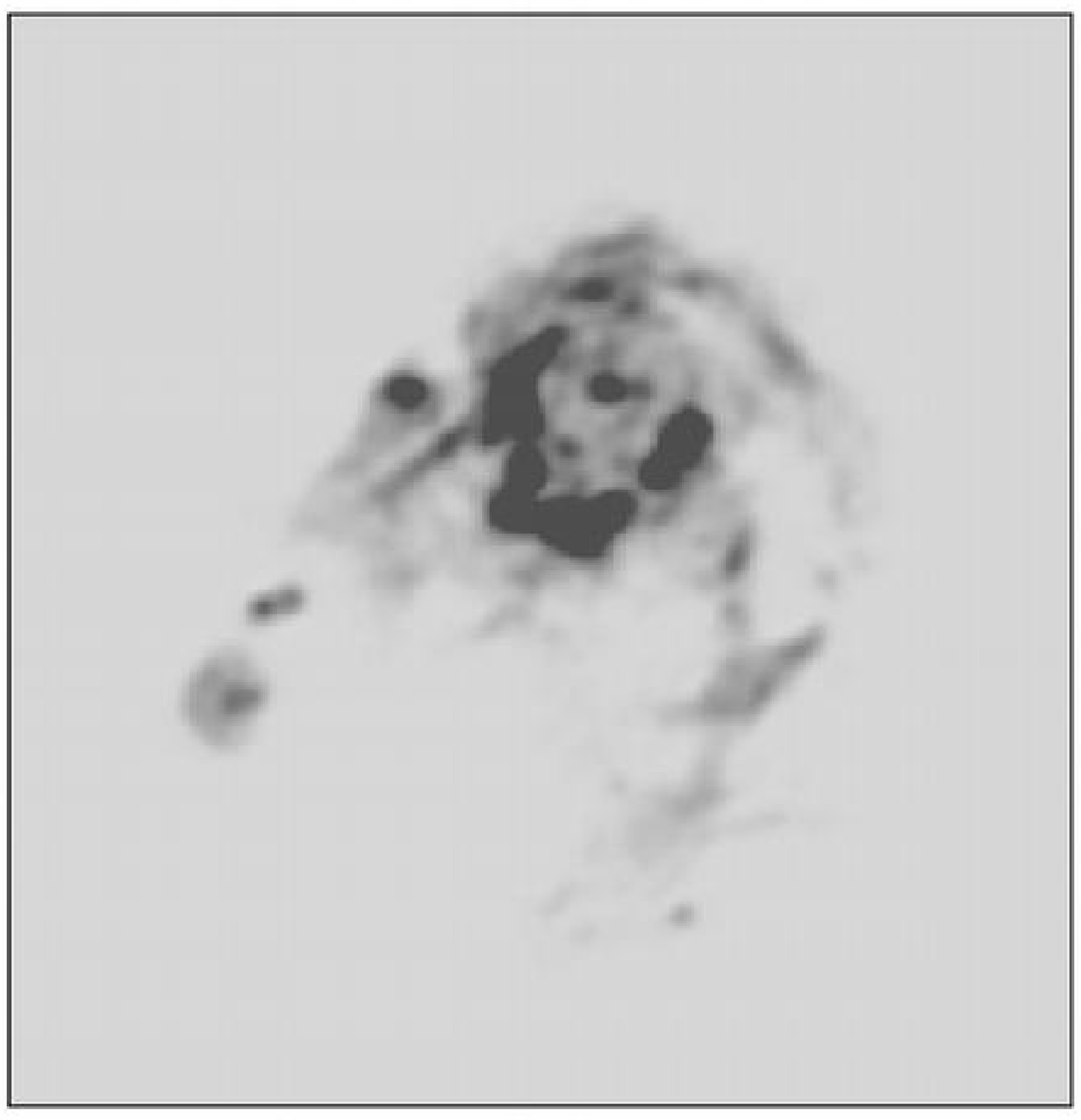}
                \includegraphics[width=8cm]{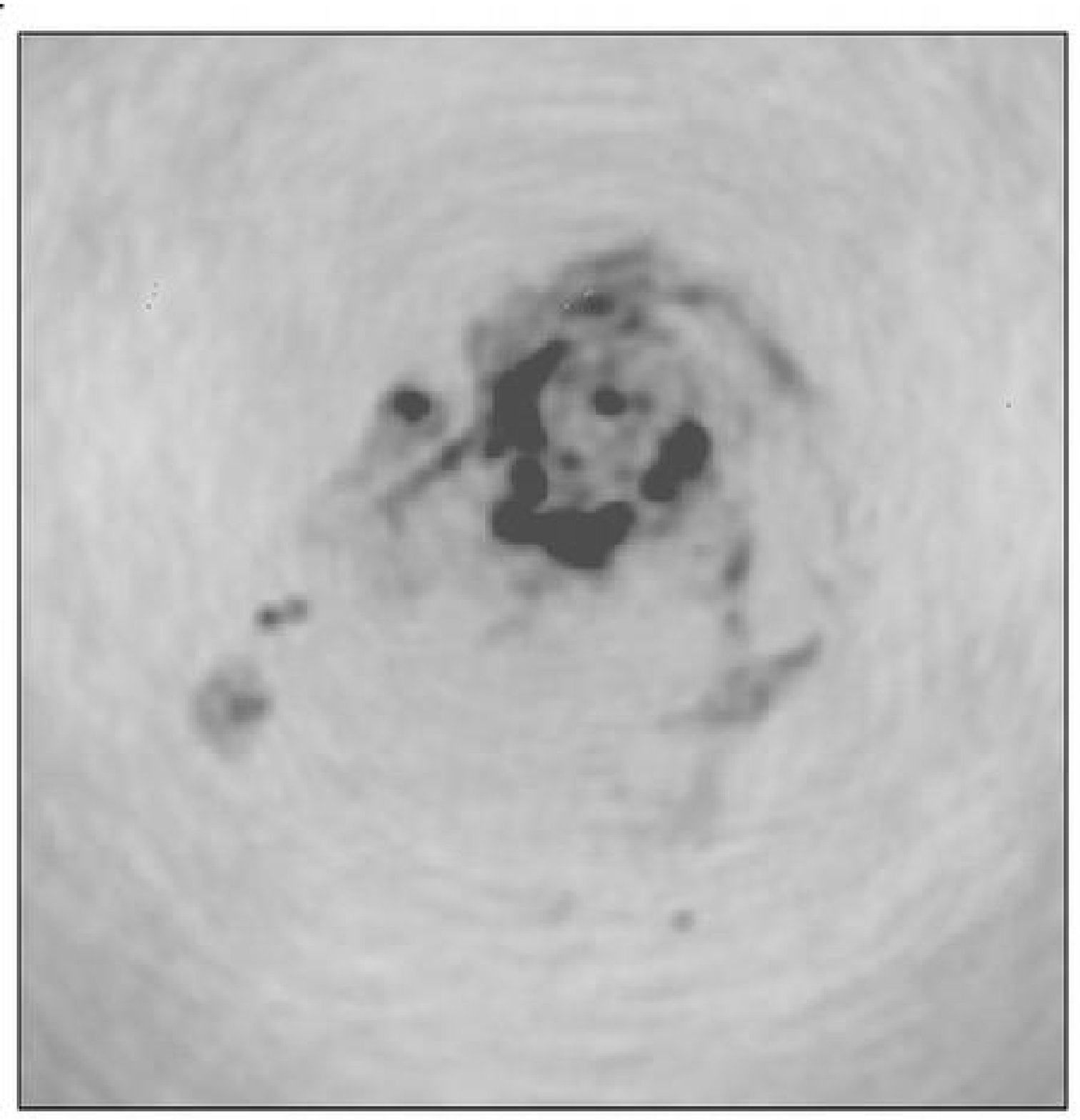}
             } 
        \hbox{
                \includegraphics[width=8cm]{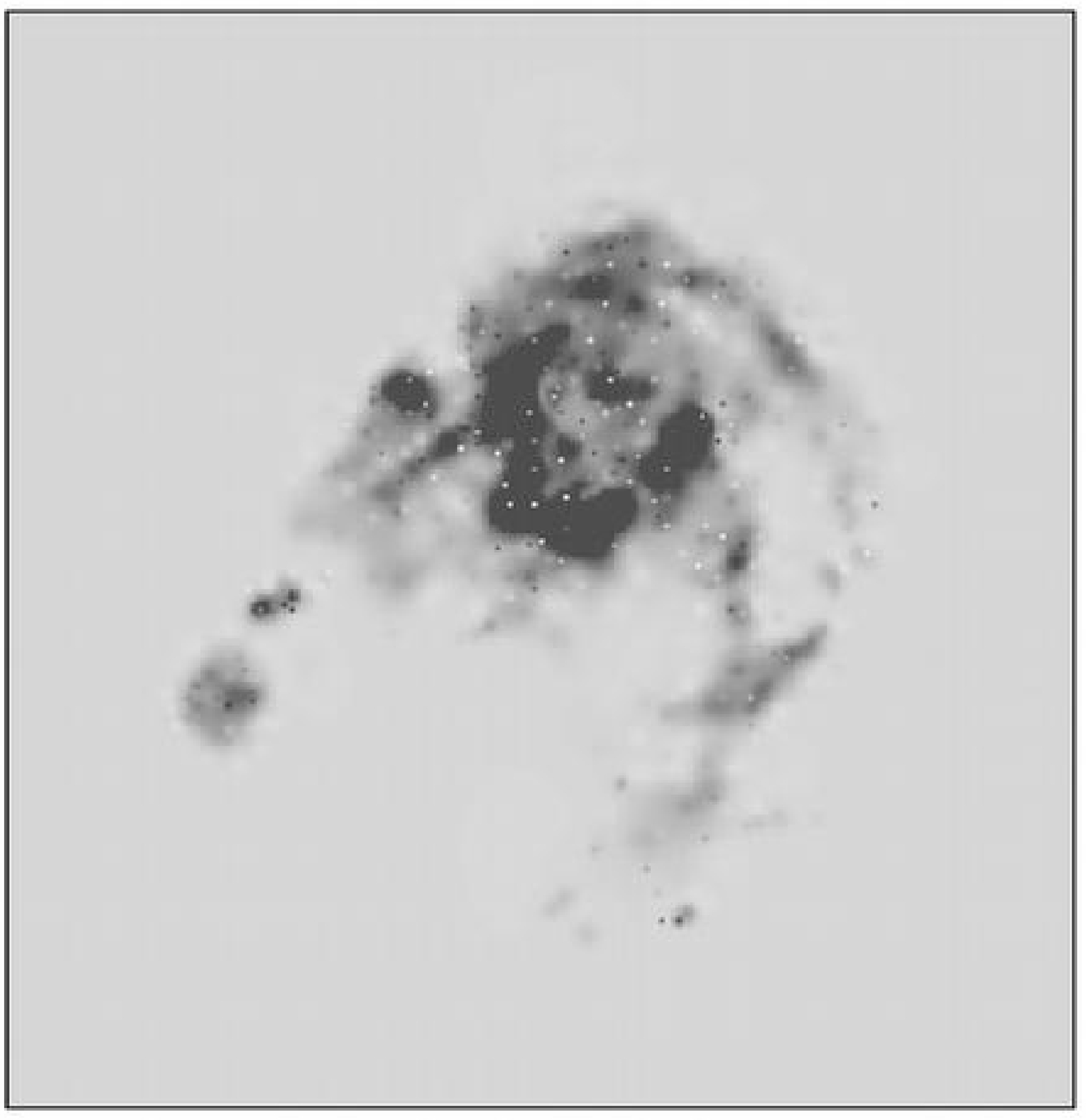}
   	        \includegraphics[width=8cm]{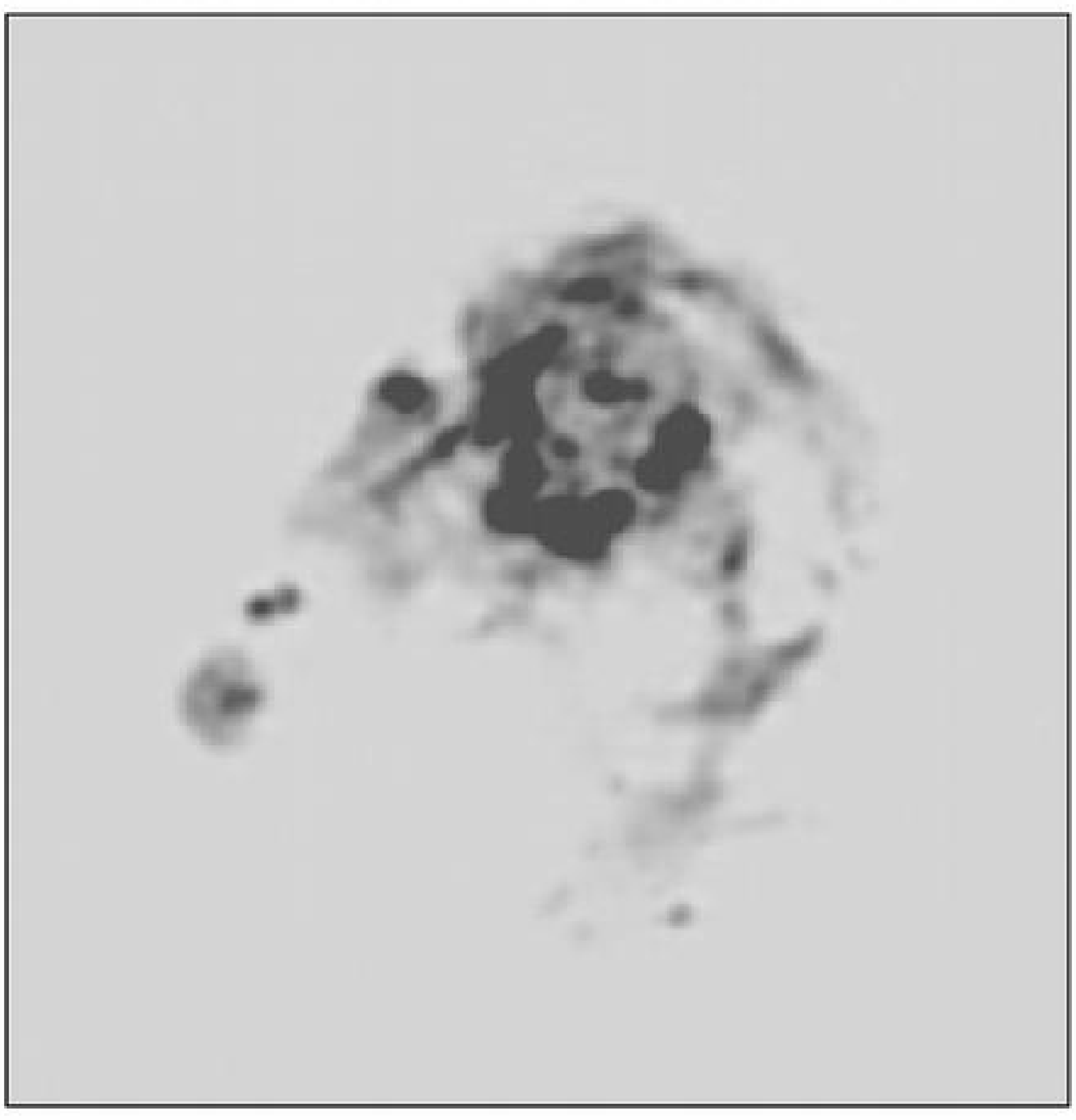}
             } 
	} 
\end{center}

   \caption{Figure showing an example of Asp reconstruction of a
   typical astronomical image.  Top left panel shows the HI image made
   with the VLA, used as the ``true image'' ($\vec{I^o}$) for the
   simulation.  The image contains $\sim 10\,000$ pixels with
   significant emission.  This image was used to simulate visibilities
   corresponding to a VLA observation.  The corresponding dirty image
   ($\vec{I^d}$), shown in the top right panel, was then deconvolved
   using the Asp-Clean algorithm.  A 800-Asp component reconstructed
   model image ($\vec{I^M}$) is shown in the lower left panel.  The
   lower right panel shows the restored Asp-model image
   ($\tens{C}\vec{I^M}+\vec{I^R}$, where $\tens{C}$ is the smoothing
   operator corresponding to the resolution element).} 
   \label{FigExM31}

\end{figure*}
The computation of the residual image, $\tens{A}^{\rm
T}\left[\vec{V}-\tens{A}\vec{I^M}\right]$, is required for the
computation of each step-size in the minimization step~\ref{OPT}.
This involves two FFTs, one gridding and one de-gridding operation.
For an $N\times N$ sized image, the cost of FFTs is of the order
$2N^2\log(N)$.  The cost of gridding and de-gridding for a $N_v$
visibility database is dominated by the $O(N_v)$ disk I/O.  For
typical sized visibility databases, the gridding costs dominate over
the cost of FFTs.  A less accurate residual computation can be done by
using a pre-gridded version of the visibilities (this will be
less accurate due to the use of gridded visibilites rather than the
raw observed visibilites).  For an Asp with an analytical expression
for its Fourier transform, the cost of the residual image computation
scales roughly as the cost of FFTs.  This form of residual computation
can be considered as the minor cycle of the Asp-Clean algorithm.
Periodically, the residual computation is done at full accuracy. 
This computation of the residuals at full accuracy and the freedom to
adjust the Aspen determined in earlier iterations ensures that the use
of gridded visibilities in the minor cycle has insignificant impact on
the final result.

\section{Examples}

The ability of the algorithm to build a model of the true image using
minimum DOF was tested by simulating an image composed of components
using the same form of Asp as is used in the deconvolution process.
It is clear that an image consisting of well separated Asp components
will pass such a test (such a case will be equivalent to a test case
for the normal CLEAN algorithm with only isolated unresolved sources).
Therefore, a test image with an overlapping set of components at
different scales (shown in the left panel of Fig.~\ref{FigExample}),
was convolved with a typical interferometric PSF to form the dirty
image (shown in the center panel of Fig.~\ref{FigExample}) and a noise
image of type $\vec{I^N}$ (RMS noise of $\approx 1$mJy) was added.
This image was deconvolved to generate the model image shown in the
right panel of Fig.~\ref{FigExample}.  Only two Asp components were
required to achieve convergence and the scale of these two components
was automatically detected.  In contrast, MS-Clean would need
more components, unless of course the selected set of scales included
precisly the two scales in the image.  This demonstrates that under
ideal conditions, where the true image is composed of {\it only} Asp
shaped components, the algorithm does not introduce extra DOF.
Although, in general where the true image is not strictly composed of
only Asp shapes, it is impossible to design an algorithm which
decomposes the image using minimum DOF, one can be optimistic that
this algorithm will use fewer DOF compared to other algorithms.  In
practice, it is indeed observed that Asp-Clean uses about an order of
magnitude fewer DOF than MS-Clean, which in turn uses few orders of
magnitude fewer DOF than scale-less decomposition.  Note that
neither of these scale sensitive algorithms attempt to combine the
components, e.g. construct a single component out of components
located at the same pixel and of similar scales, to reduce the final
number of components.

\begin{figure*}
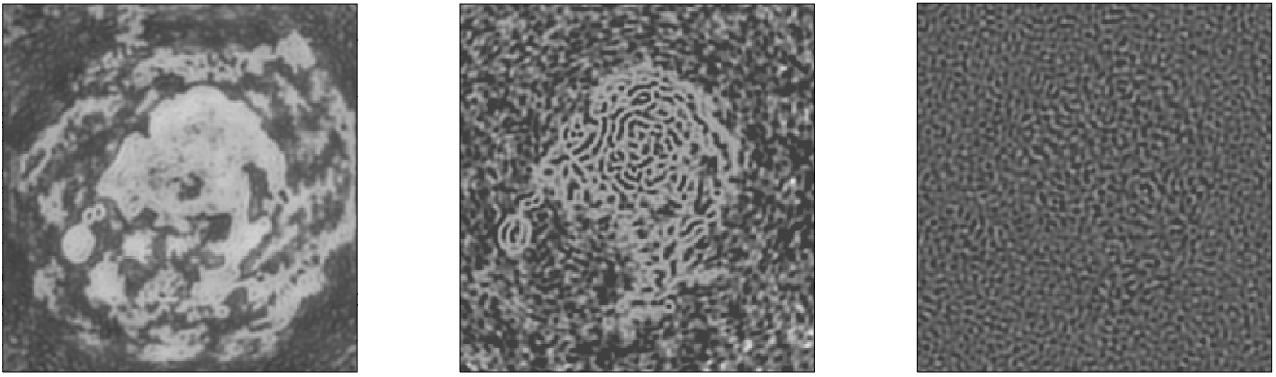

\begin{center}
\hbox{
    \includegraphics[width=6cm]{clean.res.eps}
    \includegraphics[width=6cm]{msclean.res.eps}
    \includegraphics[width=6cm]{aspclean.res.eps}
}
\end{center}
    \caption{Figure shows the comparison of the residual images
    ($\tens{B}\vec{I^M}-\vec{I^d}$) using the Cotton-Schwab Clean
    (left), MS-Clean (center) and Asp-Clean (right) algorithms.  The
    Asp model image used to compute the Asp-Clean residual was the
    same as shown in Fig.~\ref{FigExM31}.  The residuals for Asp-Clean
    are consistent with the noise and contains no correlated features
    at scales larger than the resolution element.  The residuals in
    the other two images are correlated with the large scale emission
    and are comparable in magnitude to the peak noise in the
    off-source regions.  The Cotton-Schwab Clean was run for $50\,000$
    components, MS-Clean for $20\,000$ components and Asp-Clean for
    $800$ components such that the peak residual was same for all
    cases.  MS-Clean was run with 5 scales of sizes $0, 3, 5, 10$ and
    15 pixels.}
\label{Residuals}
\end{figure*}

The top left panel of Fig.~\ref{FigExM31} shows the ``M31'' image used
as the test image in our simulation.  This was convolved with the PSF
to generate the simulated dirty image shown in the top right panel.
The resolution in the original image was $\approx2''$ while the
synthesized beam for the PSF was $\approx 2.5''$ in size.  Asp
deconvolution resulted in the Asp model image shown in the lower left
panel of Fig.~\ref{FigExM31}.  The image is composed of a number of
Aspen with the smallest being of the size of a single pixel (the
bright ``dots'' in the image).  Since the resolution element (the
synthesized beam) is larger than the pixel size, this image needs to
be smooth to the scale of the resolution element, which is shown in
the lower right panel of Fig.~\ref{FigExM31}.  However, even without
this smoothing operation, the Asp image is quite good showing all the
morphological details of the original image.  The emission,
particularly the low level extended emission, is not broken up into
small scales (as is in the case of scale-less deconvolution) and
features with strong compact emission surrounded by extended emission
are also properly reconstructed.  The lower right panel shows the
model image convolved with the estimated resolution element.
Fig.~\ref{Residuals} shows the residual images for the standard
Cotton-Schwab Clean (CS-Clean), MS-Clean and the Asp-Clean algorithms.
While the residuals for the CS-Clean and the MS-Clean are correlated
with the large scale emission in the image, the Asp-Clean residuals
are statistically consistent with the noise and have no significant
correlated features at scales larger than the resolution element.
\section{Acceleration methods}

For an $N$ Aspen set $\{P\}_n$ at iteration $n$, the dimensionality of
the search space is $M\times N$ where $M$ is the number of parameters
per Asp.  The total number of parameters monotonically increases as a
function of iteration number and step~\ref{OPT} becomes inefficient
for complex images where the total number of Aspen can be several
hundred.

\begin{figure}
  \begin{center}
    \vbox{ 
        \includegraphics[width=9cm]{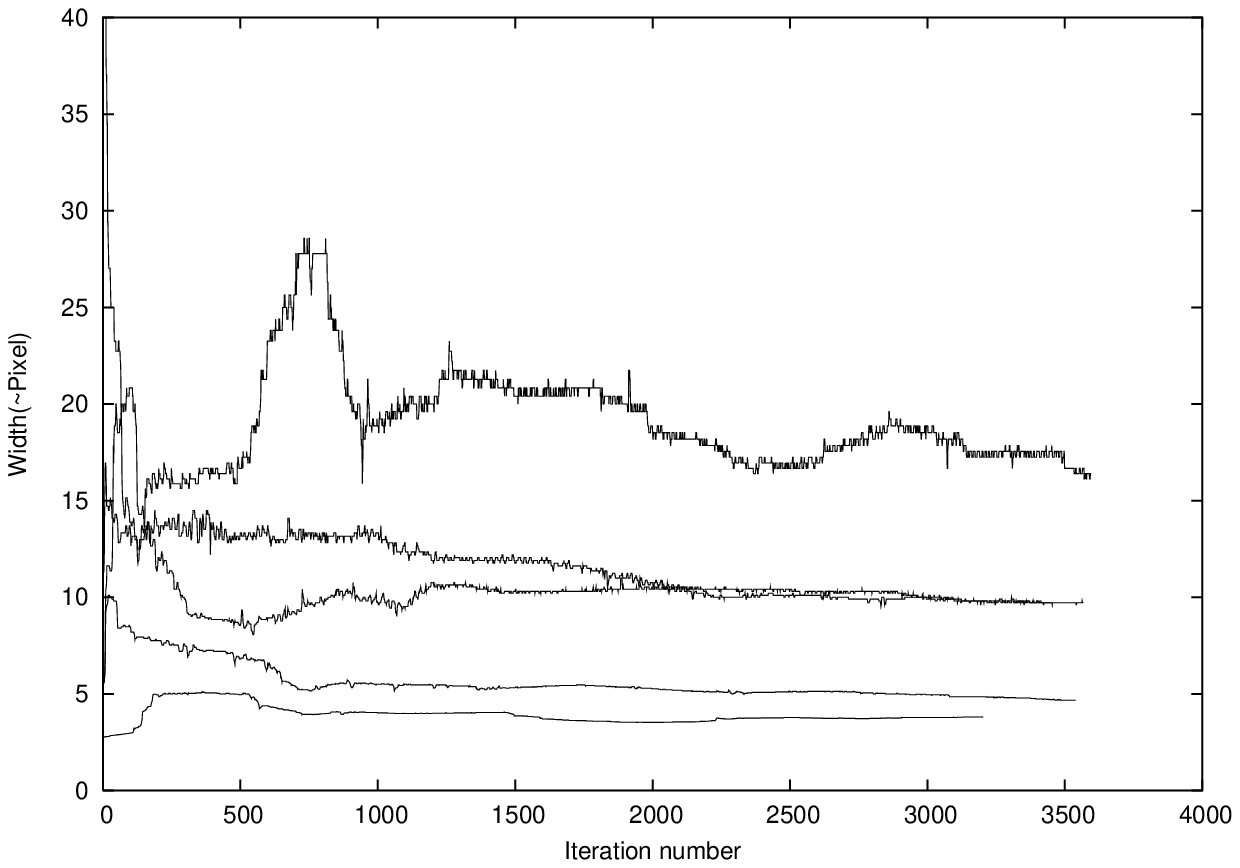}
        \includegraphics[width=9cm]{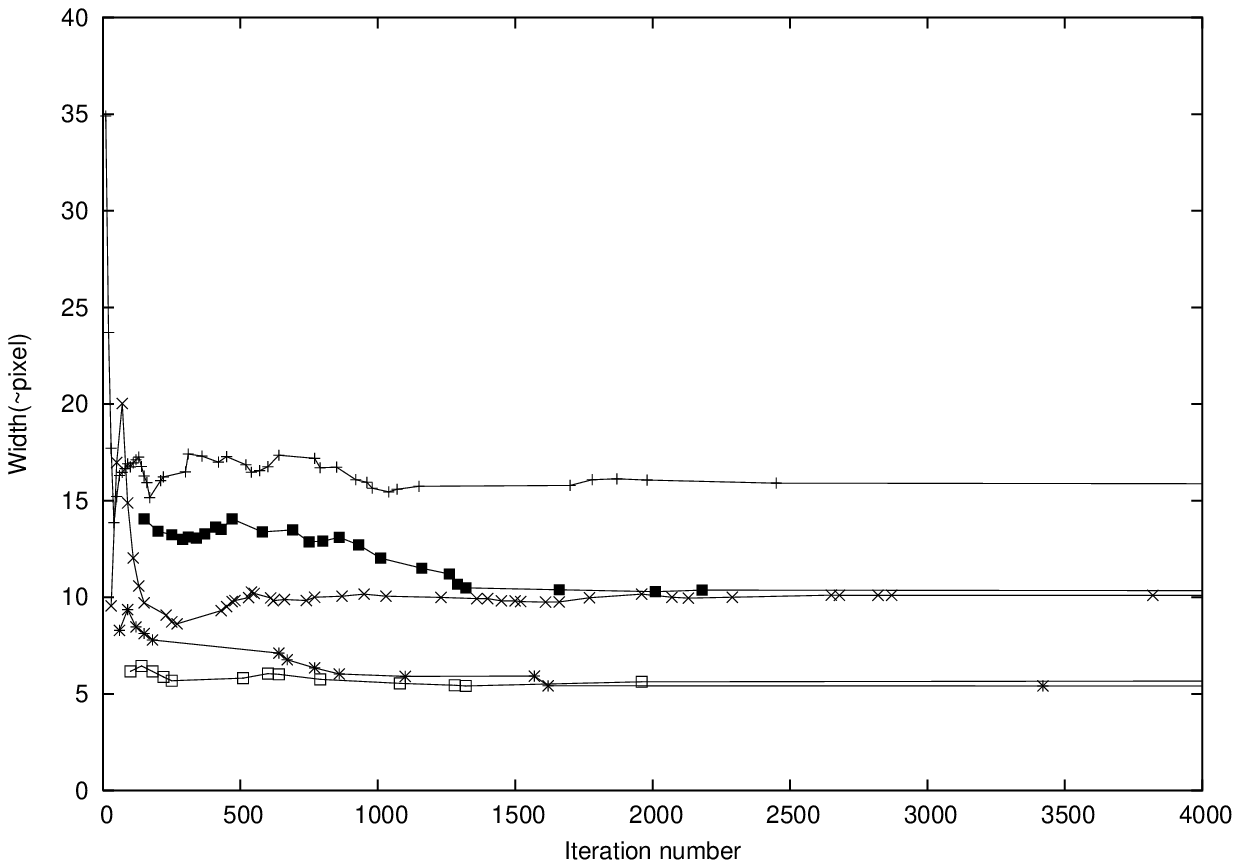}
     } 
  \end{center}

  \caption{Evolution of a few large Aspen as a function of iterations
  for a test image.  For the figure in the top panel, each Asp is kept
  in the problem after it is introduced.  It shows that the size of
  most Aspen settle down and does not change significantly at each
  iteration.  Flat portions in these curves imply insignificant step
  size in the optimization iterations.  Dropping the Aspen in
  iterations where they do not change, significantly reduces
  computational cost with minimal adverse effect on convergence.  The
  lower panel shows the evolution of the scales, after applying a
  heuristic at the beginning of each iteration to drop the Aspen which
  are unlikely to change significantly. At each iteration, only the
  Aspen marked with a symbol in this plot were retained for
  optimization for that iteration.}

  \label{FigScaleEvolution}
\end{figure}

Figure~\ref{FigScaleEvolution} shows the evolution of the scale of the
few largest Aspen for a typical test case image.  To demonstrate the
evolution of the scales, an Asp once introduced, was kept in the
problem for all successive iterations\footnote{In the spirit of
searching for an image decomposition with a minimum degree of freedom,
it is desirable to develop heuristics to completely remove Aspen which
were significant to begin with but have been rendered insignificant
due to the addition of other Aspen in successive cycles.  Furthermore,
heuristics to merge different but similar Aspen will further help in
reducing the degrees of freedom.  Work on exploring these
possibilities is in progress and will be reported in future papers.}.
The top panel of Fig.~\ref{FigScaleEvolution} shows that after initial
adjustment, the scale of most Aspen did not change significantly.  The
parts of the curves in the top panel with small derivatives correspond
to small step-sizes in the minimization algorithm.  Clearly, dropping
Aspen which are unlikely to change significantly will greatly improve
the performance with minimal adverse effect on the value of the
objective function.

\begin{figure}
  \begin{center}
    \includegraphics[width=9cm]{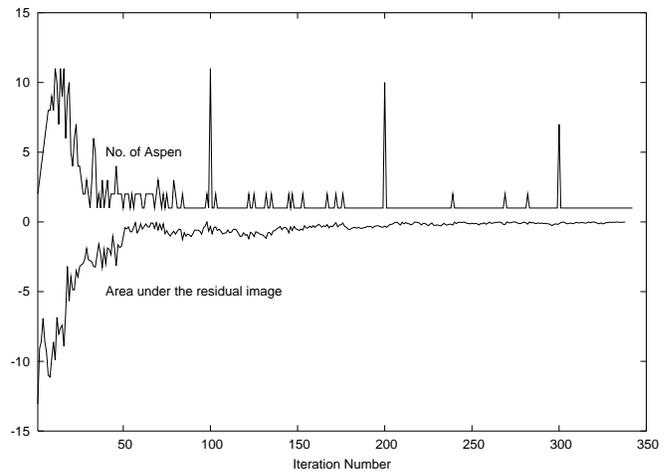}
  \end{center}
  \caption{Figure showing the number of Aspen used along with the rate
  of convergence as a function of the iteration number.  After initial
  rise in the number of Aspen, convergence was achieved by keeping
  only the latest few Aspen in the problem, effectively achieving a
  dynamic control on the dimensionality of the search space. } 
  \label{FigConvergence}
\end{figure}

For a search space with constant curvature along all axis, the update
step-size is proportional to the magnitude of the derivative along the
various axis.  Heuristically, the step-size for each Aspen in the
minimization algorithm is proportional to the length of the derivative
vector with respect to its parameters ($L_k=|\vec{\nabla_k} \chi^2|$,
where $\vec{\nabla_k}$ corresponds to the derivative operator with
respect to the parameters of the $k^{th}$ Asp only).  This suggests a
simple way of determining the active-set -- i.e. by computing $L_k$ at
the beginning of each minimization cycle and dropping Aspen for the
{\it current} cycle for which $L_k$ is below a threshold $L_o$.
Although this effectively assumes that the curvature along all axis is
constant and is the same, and will result into some mistakes, one can
recover from mistakes in later cycles since such a heuristic is
applied at the beginning of each cycle.  Assuming that the $\chi^2$
surface is well approximated by a parabola close to $\chi^2=0$, its
slope progressively decreases as convergence is approached.  $L_o$
therefore should also decrease as a function of convergence.  The area
under the residual image is indicative of the degree of convergence.
Hence, a threshold of $L_o = \lambda \sum\vec{I^R}$ was applied to
$L_k$ at the beginning of each cycle to determine the active-set.
The value of $\lambda$ controls the size of the active-set and
needs to be determined empirically based on the available computing
power and the complexity of the image - larger its value, the smaller
the size of the active-set.  Effects of the value of $\lambda$ on the
rate of convergence and the quality of reconstruction has not yet been
studied well.  The lower panel of Fig.~\ref{FigScaleEvolution} shows
the evolution of the scale of the same set of Aspen as shown in the
top panel - but after applying this heuristic.  Only the Aspen
indicated by symbols on these curves, constitute the active set at
each iteration.  Roughly speaking, Aspen which were not evolving were
automatically dropped from the problem.  The figure also shows that
the Aspen were dynamically included in the problem if it was estimated
that adjusting them would have a significant impact on the $\chi^2$.
Fig.~\ref{FigConvergence} shows the number of Aspen along with the
area under the residual image as a function of the iteration number.
For the test image used, the initial iterations correspond to the
larger Aspen.  As the process progressed towards convergence, these
large Aspen settled down, i.e. small adjustments to them had a lesser
impact on convergence, than the addition of weaker and smaller scaled
Aspen.  Consequently, the initial Aspen dropped out of the problem,
and at later times, convergence was achieved by keeping only a few
(latest) Aspen.  Effectively, this achieved a dynamic control on the
dimensionality of the search space.

Strictly speaking, the active-set depends on the structure of the PSF
and the inherent coupling of pixels in the image.  Thresholding on the
derivative vector and ignoring the curvature might result into
dropping some axis with large curvature but small slope.  Keeping them
in the problem can potentially improve the rate of convergence.
However that would require computation of the full covariance matrix,
which is prohibitively expensive.  An efficient algorithm to determine
the structure of the covariance matrix, even approximately, will be
most useful.  Work for testing some of our ideas for such an algorithm
is in progress.

\section{Summary}

Deconvolution of images with a large number of pixels with significant
emission is inherently a high dimensional problem.  The computational
load depends on the structure of the covariance matrix, which in turn
depends on the form of parameterization of the model image as well as
on the structure of the PSF.  It is possible to design efficient
algorithms to decouple the PSF and the true image for images with a
simple covariance structure (diagonal or band-diagonal matrix).  The
Pixon method is suitable for filled aperture telescopes since the PSF
has a limited support making the covariance matrix strictly band
diagonal at the worst, and the problem can be broken up into smaller
local problems.

In addition to dealing with the inherent coupling of the pixels in the
true image, interferometric imaging involves computing the coupling of
the pixels due to the PSF, often through out the image.  This has to
be done for each trial model image (or its components) in search of
the best model image.  The Pixon method is therefore unsuitable for
interferometric imaging.  MS-Clean partly takes into account the
inherent coupling of the pixels of the true image by representing it
as a collection of components at a few scales.  It ignores the
coupling between these components but takes care of the coupling due
to the PSF by pre-computing its effects.  The advantage of this
approach is that the computation of the coupling remains a
scale-and-shift operation, which is efficient, while making the
algorithm scale sensitive, albeit in a limited sense.  The
disadvantage is that it uses more DOF than necessary which leads to
similar problems as in the CLEAN algorithm (that of breaking up of
large scale emission).  Non-symmetric structures are also difficult to
accurately reconstruction.

Asp-Clean attempts to deal with these problems by explicitly solving
the problem in the hyper-space of the Aspen parameters.  The true
image is modeled in the continuous space of Aspen.  Since it is
easy to parameterize the Aspen such that non-symmetric Aspen are also
allowed, this algorithm deals with the problem of non-symmetric
structures well, and results in a model image with the least DOF as
compared to other algorithms for interferometric imaging. E.g., a
purely elliptical gaussian shaped feature will be broken up into
several symmetric components in the case of MS-Clean, while Asp-Clean
can represent it with a single component.  However the computation of
the coupling is inefficient making the search for the parameters
expensive.  Since not all parameters continue to change throughout the
search process, we have implemented heuristics to determine the
active-set of parameters and dynamically limit the dimensionality of
the search space.  This improves the performance by an order of
magnitude or more.  However the overall runtime is still about a
factor of three more than MS-Clean.  Asp-Clean approach
represents relaxation of compute-saving assumptions built into Clean
and MS-Clean approaches.  It is therefore inherently more compute
intensive than other deconvolution algorithms.  Although it is
certainly useful to develop heuristics which will minimize redundant
intermediate computations, the Asp decomposition is limited by the
compute load for the search for the Asp parameters which in turn
scales strongly with the size of the active-set at each iteration.
Since the size of the active-set of Aspen is roughly a measure of the
degree to which the non-orthogonality of the Aspen space (coupling
between the Aspen) is incorporated in the algorithm, it is fair to
expect it to impact the rate of convergence significantly.  Faster
convergence and possibly better reconstruction can be achieved with
more computing power.  Asp-Clean therefore scales better with the CPU
speed, compared to other scale sensitive algorithms, and therefore has
the potential of benefiting more from the Moore's law of CPU speeds.

Currently we are using the standard routines for the search algorithm
(conjugate gradient method and its variants or the Levenberg-Marquardt
algorithm; see \citet{GSL} for details).  Fine tuning this
minimization specifically for this problem will further improve the
runtime performance.  We use gaussians as the functional form for Asp.
The scale of such Aspen is controlled by the
full-width-at-half-maximum of the gaussian function.  However there is
no limit on the shape of the Aspen.  More sophisticated functional
forms which allow independent control on the shape as well the scale
of the Asp will further reduce the number of components needed.  Also,
improving the heuristics to determine the active-set of Aspen, and
making the heuristic computation itself efficient will be a worthwhile
future direction of work.

Finally, we note that most of the Aspen with scale significantly
larger than zero are found in the initial few hundred iterations,
where the extra computational cost for Asp-Clean is well justified.
The computational cost does not reduce for zero (or close to zero)
scale Aspen.  It may be most effective to develop heuristics to switch
to other scale-less algorithms at late times which are more efficient
for zero scale Aspen.  Work in all these directions is in progress.
Report on this on-going work and the impact of such algorithms on
the imaging performance of future interferometric telescopes and
imaging modes like mosaicking will be the subject matter of future
papers.

The Asp-Clean algorithm as described in this paper is implemented
as a Glish client in AIPS++ and can be run via a Glish script.  Work
is underway to incorporate this as one of the many available deconvolution
algorithms in the standard AIPS++ interferometric imaging tool.


\begin{acknowledgements}
We thank the AIPS++ Project for an excellent prototyping and
development environment for research in algorithm development.  We
also thank the TGIF group in Socorro for many discussions on this and
related subjects.  SB thanks Urvashi~R.V., S.~Upreti and D.~Oberoi for
numerous discussions.  TJC thanks Rick Puetter and Amos Yahil for
numerous discussions.  We thank the anonymous referee for helpful
comments.  All of this work was done on computers running the
GNU/Linux operating system and we thank all the numerous contributers
to this software.
\end{acknowledgements}

\bibliographystyle{aa}
\bibliography{asp}

\end{document}